\documentclass[12pt]{article}

\usepackage{epsfig,epsf}

\begin{document}

\def\beq{\begin{equation}}

\def\eeq{\end{equation}}
\def\eq#1{{Eq.~(\ref{#1})}}

\def\fig#1{{Fig.~\ref{#1}}}
             
\newcommand{\bas}{\bar{\alpha}_S}

\newcommand{\as}{\alpha_S} 

\newcommand{\bra}[1]{\langle #1 |}

\newcommand{\ket}[1]{|#1\rangle}

\newcommand{\bracket}[2]{\langle #1|#2\rangle}

\newcommand{\intp}[1]{\int \frac{d^4 #1}{(2\pi)^4}}

\newcommand{\mn}{{\mu\nu}}

\newcommand{\tr}{{\rm tr}}

\newcommand{\Tr}{{\rm Tr}}

\newcommand{\T} {\mbox{T}}

\newcommand{\braket}[2]{\langle #1|#2\rangle}

\newcommand{\ab}{\bar{\alpha}_S}

\setcounter{secnumdepth}{7}

\setcounter{tocdepth}{7}

\parskip=\itemsep               %?

\setlength{\itemsep}{0pt}       %?

\setlength{\partopsep}{0pt}     %?

\setlength{\topsep}{0pt}        %?

%---layout fuer eine dina4 seite-------------------

\setlength{\textheight}{22cm}

\setlength{\textwidth}{174mm}

\setlength{\topmargin}{-1.5cm}

%\input psfig

%%%%%%%%%%%%%%%%%%%%%%%%%%%%%%%%%%%%%%%%%%%%%%%%%%%%%%%%%%%%%%%

%\renewcommand{\thefootnote}{\fnsymbol{footnote}}

\newcommand{\beqar}[1]{\begin{eqnarray}\label{#1}}

\newcommand{\eeqar}{\end{eqnarray}}

\newcommand{\m}{\marginpar{*}}

\newcommand{\lash}[1]{\not\! #1 \,}

\newcommand{\nn}{\nonumber}

\newcommand{\D}{\partial}

\newcommand{\h}{\frac{1}{2}}

\newcommand{\g}{{\rm g}}

\newcommand{\el}{{\cal L}}

\newcommand{\A}{{\cal A}}

\newcommand{\Ka}{{\cal K}}

\newcommand{\al}{\alpha}

\newcommand{\be}{\beta}

\newcommand{\ep}{\varepsilon}

\newcommand{\ga}{\gamma}

\newcommand{\de}{\delta}

\newcommand{\De}{\Delta}

\newcommand{\et}{\eta}

\newcommand{\ka}{\vec{\kappa}}

\newcommand{\la}{\lambda}

\newcommand{\ph}{\varphi}

\newcommand{\si}{\sigma}

\newcommand{\ro}{\varrho}

\newcommand{\Ga}{\Gamma} 

\newcommand{\om}{\omega}

\newcommand{\La}{\Lambda}  

\newcommand{\tG}{\tilde{G}}

\renewcommand{\theequation}{\thesection.\arabic{equation}}

%%%%%%%%%%%%%%%%%%%%%%%%%%%%%%%%%%%%%%%%%%%%%%%%%%%%%%%%%%%%%%%%%

% ABBREVIATED JOURNAL NAMES  

%

\def\ap#1#2#3{     {\it Ann. Phys. (NY) }{\bf #1} (19#2) #3}

\def\arnps#1#2#3{  {\it Ann. Rev. Nucl. Part. Sci. }{\bf #1} (19#2) #3}

\def\npb#1#2#3{    {\it Nucl. Phys. }{\bf B#1} (19#2) #3}

\def\plb#1#2#3{    {\it Phys. Lett. }{\bf B#1} (19#2) #3}

\def\prd#1#2#3{    {\it Phys. Rev. }{\bf D#1} (19#2) #3}

\def\prep#1#2#3{   {\it Phys. Rep. }{\bf #1} (19#2) #3}

\def\prl#1#2#3{    {\it Phys. Rev. Lett. }{\bf #1} (19#2) #3}

\def\ptp#1#2#3{    {\it Prog. Theor. Phys. }{\bf #1} (19#2) #3}

\def\rmp#1#2#3{    {\it Rev. Mod. Phys. }{\bf #1} (19#2) #3}

\def\zpc#1#2#3{    {\it Z. Phys. }{\bf C#1} (19#2) #3}

\def\mpla#1#2#3{   {\it Mod. Phys. Lett. }{\bf A#1} (19#2) #3}

\def\nc#1#2#3{     {\it Nuovo Cim. }{\bf #1} (19#2) #3}

\def\yf#1#2#3{     {\it Yad. Fiz. }{\bf #1} (19#2) #3}

\def\sjnp#1#2#3{   {\it Sov. J. Nucl. Phys. }{\bf #1} (19#2) #3}

\def\jetp#1#2#3{   {\it Sov. Phys. }{JETP }{\bf #1} (19#2) #3}

\def\jetpl#1#2#3{  {\it JETP Lett. }{\bf #1} (19#2) #3}

%%%%%%%%% notice the parenthesys is only on one side

\def\ppsjnp#1#2#3{ {\it (Sov. J. Nucl. Phys. }{\bf #1} (19#2) #3}

\def\ppjetp#1#2#3{ {\it (Sov. Phys. JETP }{\bf #1} (19#2) #3}

\def\ppjetpl#1#2#3{{\it (JETP Lett. }{\bf #1} (19#2) #3} 

\def\zetf#1#2#3{   {\it Zh. ETF }{\bf #1}(19#2) #3}

\def\cmp#1#2#3{    {\it Comm. Math. Phys. }{\bf #1} (19#2) #3}

\def\cpc#1#2#3{    {\it Comp. Phys. Commun. }{\bf #1} (19#2) #3}

\def\dis#1#2{      {\it Dissertation, }{\sf #1 } 19#2}

\def\dip#1#2#3{    {\it Diplomarbeit, }{\sf #1 #2} 19#3 }

\def\ib#1#2#3{     {\it ibid. }{\bf #1} (19#2) #3}

\def\jpg#1#2#3{        {\it J. Phys}. {\bf G#1}#2#3}  

%

%%%%%%%%%%%%%%%%%%%%%%%%%%%%%%%%%%%%%%%%%%%%%%%%%%%%%%%%%%%%%%%%%%%%%

%

%\renewcommand{\thefigure}{{\protect\bf\arabic{figure}}}

\def\thefootnote{\fnsymbol{footnote}} 

%

%    

%

%\begin{titlepage}

\noindent

\begin{flushright}

\parbox[t]{10em}{

 \today 
}
\end{flushright}

\vspace{1cm}

\begin{center}

{\LARGE  \bf  Balitsky's hierarchy from Mueller's dipole}\\
{\LARGE  \bf  model and more about target correlations.}\\

\vskip1cm

{\large \bf ~E. ~Levin ${}^{a \,\ddagger}$ 
\footnotetext{${}^{\ddagger}$ \,\,Email:
leving@post.tau.ac.il, levin@mail.desy.de.}  
and ~M. ~Lublinsky ${}^{b \,\star}$ 

\footnotetext{${}^{\,\star}$ \,\,Email:
lublinsky@phys.uconn.edu, lublinm@mail.desy.de }}

\vskip1cm

{\it ${}^{a)}$\,\,\, HEP Department}\\
{\it School of Physics and Astronomy}\\
{\it Raymond and Beverly Sackler Faculty of Exact Science}\\
{\it Tel Aviv University, Tel Aviv, 69978, Israel}\\
\vskip0.3cm
{\it ${}^{b)}$\,\,\, Physics Department} \\
{\it University of Connecticut} \\
{\it U-3046, 2152 Hillside Rd., Storrs} \\
{\it CT-06269, USA}\\
\vskip0.3cm

\end{center}  

\bigskip

\begin{abstract}        
High energy scattering formulated as a classical branching process
is considered within the 
framework of the
QCD dipole model.
Starting from   Mueller's generating functional, we derive the high 
energy  evolution law for the scattering amplitude. The amplitude's 
evolution is given by an infinite hierarchy of linear equations
equivalent to the Balitsky's chain reduced to dipole operators. This
new derivation of the hierarchy is the central result of the paper.
We also comment about target correlations which prevent the hierarchy
from  being expressed as  the  Balitsky-Kovchegov equation in closed form.    
\end{abstract}

\newpage

%*********************************************************************************  

%*********************************************************************************  

\def\thefootnote{\arabic{footnote}} 

\section{Introduction}

The high energy scattering in QCD can be most efficiently addressed
in terms of colour dipole degrees of freedom. In this approach, originally
proposed by Mueller  \cite{MUUN}, one considers a fast moving projectile
as a  bunch of dipoles created via a classical branching process. High
energy evolution of the projectile wavefunction can be then found from the
QCD generating functional  \cite{MUUN}, which obeys a  linear functional
evolution equation \cite{LL}. Though the evolution of the projectile 
wavefunction is independent of target, the interaction amplitude  depends on it. 

The energy evolution of the interaction amplitude will be the central object 
studied in the present paper. It was shown by Kovchegov \cite{K},
 that if one assumes that all the 
dipoles produced via evolution interact with the target independently
of each other,  then the scattering amplitude obeys a nonlinear
evolution equation presently referred to as the Balitsky-Kovchegov (BK) 
equation \cite{B,K}. The BK equation has been the core  of numerous 
analytical \cite{BKT} and numerical \cite{BKN} studies. 
It has  also been  applied for phenomenology \cite{LUB}.

It was realised in Ref. \cite{LL},  that the assumption regarding independent 
interactions can be incorrect  for many targets, and in particular for 
realistic nuclei which are not very dense. Target correlations are needed 
in order  to account for more realistic structure of targets. 
It was shown in Ref. \cite{LL} that the amplitude obeys 
a  linear functional evolution equation,  which is the most appropriate tool
for the  introduction of  target correlations. 
In general, the linear functional equation for the amplitude with
target correlations cannot be represented as an ordinary equation and preserves its functional 
form. 
Our goal here is to show that the linear functional 
evolution equation for  the interaction amplitude
for a generic target with correlations 
is equivalent to an infinite chain of linear  hierarchy equations.
This chain is 
the hierarchy of Balitsky \cite{B} reduced to dipole operators
only \footnote{The true Balitsky's hierarchy, as well as its equivalent, 
the JIMWLK equation \cite{ELTHEORY}, also
contain  non-dipole operators, which account for
$N_c$ corrections. In the present paper we have nothing to say about $N_c$
corrections which will be systematically ignored.}. Thus 
we clarify the connection between the Wilson loop  approach of Balitsky and  the Mueller dipole 
model.    
We also demonstrate that the large $N_c$ limit, which is in the foundation
of the dipole model, is insufficient to guarantee a close form equation
for the scattering amplitude. This new derivation of the Balitsky`s 
chain should be view as our main result. 

We also demonstrate a possibility to rigorously derive an evolution
for the amplitude without restricting ourselves 
to a single dipole as a specific choice of projectile. Our results 
are valid for any projectile whose  evolution is not affected by high density
effects.

At the end of the paper we discuss an ansatz for target
correlations which allows one  to obtain a certain generalisation of the BK 
equation. This part is very much along the lines of Ref. \cite{LL}.
Similar result has been recently reported by Janik \cite{Janik}.

One of our central goals in this letter 
is to draw more attention to Ref.\cite{LL} in which the 
linear functional formalism was developed  as well as 
many other results appearing in this reference.

\section{Balitsky's chain from the QCD generating functional}

We first consider a generic fast moving 
projectile whose wave function can be expanded in a 
dipole basis. Contrary to many previous studies we do not
restrict ourselves to a single dipole as a projectile.
We further assume that non-linear effects associated with high
dipole densities in the projectile wave function can be ignored.
This is a strong assumption which eliminates  effects which might be associated with pomeron loops, so called
enhanced diagrams. 

Let us define a  probability density $P_n$ to 
find  $n$ dipoles with coordinates $r_1, b_1$, $r_2, b_2$, $\dots r_i, 
b_i$, $\dots\,r_n, b_n$ 
and  rapidity $Y$ in the projectile wave function. 
$r_i$ and $b_i$ denote the  dipole's size
and impact parameter, both are two dimensional vectors.  
We define $P_n$ as a dimensionfull quantity.
The equation for $P_n$ obeys the classical branching process:
\begin{eqnarray}\label{LEQPF}
\,\,\frac{\partial\,P_n\left(Y\,-\,Y_0;\,r_1, b_1,\,r_2, b_2,
\dots r_n, b_n \right)}{ 
\bar{\alpha}_s\,\partial\, Y}\,=\,-\,
\sum^n_{i=1}\,\omega(r_i) \,
P_n\left(Y\,-\,Y_0;\,r_1, b_1,\,r_2, b_2, \,\dots r_n, b_n
\right)\nonumber \\
+\,\sum^{n-1}_{i=1} \,\frac{(r_i\,+\,r_n)^2}{(2\,\pi)\,r^2_i\,r^2_n}\,
P_{n - 1}\left(Y\,-\,Y_0;\,r_1, b_1,\,r_2, b_2,
\dots  (r_i \,+\, r_n), b_{in},\dots r_{n-1}, b_n
 \right)
\end{eqnarray}
with $b_{in}\,=\,b_i\,+\,r_n/2\,=\,b_n\,-\,r_i/2$
being imposed via $\delta$-functions.
Two terms of Eq.(\ref{LEQPF}) have a 
very simple meaning: the first one describes
the decrease in probability to find $n$ dipoles due to a decay of one dipole
into two  of arbitrary sizes.  This probability is equal to
$$
 \bar{\alpha}_s \,\,\omega(r_i)\,\,=\,\,\frac{ 
\bar{\alpha}_s}{2\,\pi}\,\int_\rho 
\,\frac{r_i^2}{(r_i\,-\,r')^2\,r'^2}\,d^2 r'\,\,=\,\,
\bar{\alpha}_s \,\,\ln(r_i^2/\rho^2)
$$
with $\rho$ being some infrared cutoff, and $\bar\as\,=\,\as\,N_c/\pi$.
The second term shows the increase in probability to find $n$  dipoles
due to a creation of a new dipole from  $n -1$ dipoles with probability
$$ \frac{\bar{\alpha}_s}{2\,\pi} \,\,\frac{(r_1 \,\,+\,\, 
r_2)^2}{r_1^{2}\,r_2^2}\,\,.
$$
Eq. (\ref{LEQPF}) has to be supplemented by initial conditions at $Y=Y_0$,
specifying a dipole distribution in the projectile. In writing the equation
 (\ref{LEQPF}) we explicitly used our assumption that there are
no nonlinear effects
in the projectile. Otherwise, we would need to include dipole recombination
processes.

The hierarchy (\ref{LEQPF}) can be resolved by
introducing a generating functional $Z$
\beq \label{LD1}
Z\left(Y\,-\,Y_0;\,[u] \right)\,\,\equiv\,\,\sum_{n=1}\,\int\,\,
P_n\left(Y\,-\,Y_0;\,r_1, b_1,\,r_2, b_2, \dots ,r_i, b_i, \dots ,r_n, b_n
 \right) \,\,
\prod^{n}_{i=1}\,u(r_i, b_i) \,d^2\,r_i\,d^2\,b_i
\eeq 
where $u(r_i, b_i) \equiv u_i $ is an arbitrary function of $r_i$ and $b_i$. 
It  follows immediately from (\ref{LEQPF})
that the functional (\ref{LD1}) obeys the condition:
at $u\,=\,1$ 
\beq \label{LDIN2} 
Z\left(Y\,-\,Y_0;\,[u=1]\right)\,\,=\,\,1\,.
\eeq
The physical meaning of (\ref{LDIN2}) is that the sum over
all probabilities is one.

Multiplying \eq{LEQPF} by the product $\prod^n_{i=1}\,u_i$ 
and integrating over all $r_i$ and $b_i$,  we obtain the 
following linear equation for the generating functional:
\beq \label{LEQZF}
\,\frac{\partial \,Z}{
\bar{\alpha}_s\,\partial \,Y}\,\,=\,\,-\,\,
\int\,d^2 r\,d^2 b\,\,V_{1\rightarrow 1}(r,\,b,\,[u])
\,\, Z\,\,
+\,\,\int\,\,d^2 \,r \,d^2\,r' \,d^2 b\,\,V_{1\rightarrow 2}
(r,\,r',\,b,\,[u])
\,\, Z\,.
\eeq
with the definitions
\beq \label{V2} 
V_{1 \rightarrow 1}(r,\, b,\,[u])\,\,=\,\,
\bar{\alpha}_s \,\,\omega(r) \,\,u(r,\,b)\,\,\frac{\delta}{\delta u(r,b)}
\eeq
and
\beq \label{V1}
V_{1 \rightarrow 2}(r,\,r',\,b,\,[u])\,\,
=\,\,\frac{ \bar{\alpha}_s}{2\,\pi} 
\,\,\frac{r^2}{r'^2\,(r - r')^2}\,\,
\,u(r', \,b\,+\,\frac{r\,-\,r'}{2})\,\,u(r \,-\,r', \,b\,-\,\frac{r'}{2})\,\,
\frac{\delta}{\delta u(r, \,b)}\,.
\eeq 
The functional derivative with respect to $u(r,b)$,  plays the  role 
of an  annihilation operator for a dipole of the size $r$,  at the impact 
parameter $b$. 
The multiplication by $u(r,b)$ corresponds to
a creation operator for this dipole.
Eq. (\ref{LEQZF}) has an extra $b$ integration compared to the one 
first derived in Ref. \cite{LL}. 
Motivated by the fact that evolution kernels do not depend on the impact
parameter, we ignored the $b$-dependence in Ref. \cite{LL}  in order 
to simplify the presentation. This procedure also corresponds to an 
approximation in which the impact parameter is considered to be much larger
than  any dipole size. The correct analysis has to preserve the 
true kinematics and this amounts to tracing the $b$-dependence which is 
done in Eq.(\ref{LEQZF}).
Eq.(\ref{LEQZF}) can be also obtained
starting from the JIMWLK equation \cite{Janik}.

The $n$-dipole densities in the projectile 
$\rho^p_n(r_1, b_1,\ldots\,,r_n, b_n)$
 are defined as follows:
$$
\rho^p_n(r_1, b_1\,
\ldots\,,r_n, b_n; Y\,-\,Y_0)\,=\,\frac{1}{n!}\,\prod^n_{i =1}
\,\frac{\delta}{\delta
u_i } \,Z\left(Y\,-\,Y_0;\,[u] \right)|_{u=1}
$$
One can recast the  chain of equations  (\ref{LEQPF}) into hierarchy
for $\rho^p_n$:
$$
\frac{\partial \,\rho^p_n(r_1, b_1\,\ldots\,,r_n, b_n)}{ 
\bar{\alpha}_s\,\partial\,Y}\,\,=\,\,\left(
-\,\sum_{i=1}^n
 \,\,\omega(r_i)\,\,\rho^p_n(r_1, b_1\,\ldots\,,r_n, b_n)\,\,+
\right.
$$
\beq \label{DFZ}
\left.
2\,\sum_{i=1}^n\,
\int\,\frac{d^2\,r'}{2\,\pi}\,
\frac{r'^2}{r^2_i\,(r_i\,-\,r')^2}\,
\rho^p_n(\ldots\,r', b_i-r'/2\dots)\,
+\,\sum_{i=1}^{n-1}\,\frac{(r_i + r_n)^2}
{(2\,\pi)\,r^2_i\,r^2_n}\,
\rho^p_{n-1}(\ldots\,(r_i\,+\,r_n), b_{in}\dots)\right).
\eeq
The evolution for $\rho^p_1$ was considered by Mueller in his original work
\cite{MUUN}.
It obeys the linear BFKL equation \cite{BFKL}.  The $n$-dipole densities
with $n>1$, are necessary generated through the evolution even if the projectile
is a single dipole. For this specific choice of the projectile, our
linear equation, say, for $\rho^p_2$ can be identically reformulated
as a nonlinear with respect to $\rho^p_1$. This form of the equation
for $\rho_2^p$ was studied in Refs. \cite{MUUN, P}. We would like to stress,
however, that Eq. (\ref{DFZ}) is more general, as it does not depend on any
specific choice of projectile. Our
approach cannot be rigorously applied to a hadron or nucleus as a projectile, 
as the latter need to account for  high dipole densities, in spite of this  
we view our finding as  a first  important step  towards  a  symmetric 
description of  target and projectile.

So far we have discussed the evolution of the projectile only. Our goal
is to compute the total scattering amplitude.
Following Ref. \cite{K} the amplitude $N$ is defined  
\beq \label{AMPZ}
 N(Y) \,\,=\,\,-\,\, \sum^{\infty}_{n=1}\,\,(-1)^n\,
\int\,\gamma_n(r_1, b_1\ldots \,,r_n, b_n;Y_0)\,\,
\rho^p_n(r_1,b_1\,\ldots\,,r_n, b_n; Y\,-\,Y_0)
\,\,\prod^n_{i =1}\,
d^2 r_i \, d^2 b_i \,.
\eeq 
The amplitude for simultaneous 
scattering of $n$ dipoles off the target  is denoted by $\gamma_n$.
It has to be specified at the lowest rapidity ($Y_0$).
 $\gamma_n$ can be expressed through $\rho^t_n$, the
dipole densities in the target:
\beq\label{target}
\gamma_n(r_1, \,b_1,\ldots\,,r_n, b_n; Y_0)\,\,\,=\,\,\,\,\,\,\,\,\,\,
\,\,\,\,\,\,\,\,\,\,\,\,\,\,\,\,\,\,\,\,\,\,\,\,\,\,\,
\,\,\,\,\,\,\,\,\,\,\,\,\,\,\,\,\,\,\,\,\,\,\,\,\,\,\,
\,\,\,\,\,\,\,\,\,\,\,\,\,\,\,\,\,\,\,\,\,\,\,\,\,\,\,\,\,
\,\,\,\,\,\,\,\,\,\,\,\,\,\,\,\,\,\,\,\,\,\,\,\,\,\,\,
\,\,\,\,\,\,\,\,\,\,\,\,\,\,\,\,\,\,
\,\,\,\,\,\,\,\,\,\,\,\,\,\,\,\,\,\,\,\,\,\,\,\,\,\,\,
\eeq
$$
\sum^\infty_{m=1}\int \, 
\si^{n\,m}(r_1,\, b_1,\dots r_n,\,b_n\, |\,
\bar r_1,\, \bar b_1,\,\dots,\,\bar r_m,\,\bar b_m)\,\,
\rho^t_m(\bar r_1,\,\bar b_1\,\ldots\,,\bar r_m\,\bar b_m; Y_0)\,\,
\,d^2\,\bar r_1\,d^2\,\bar b_1\dots\,d^2\,\bar r_m\,d^2\,\bar b_m,
$$
with $\si^{n\, m}$ being the properly  normalised amplitude 
for $n$ dipoles  scattering off  $m$ dipoles, 
 at the two gluon exchange level for each pair of dipoles. 

While the generating functional $Z$ contains
information about the projectile only,
the target enters the amplitude  through the functions 
$\gamma_n$
($\rho^t_m$). Substituting (\ref{target})
into (\ref{AMPZ}) we obtain:
\beq \label{AMPZ1}
 N(Y) \,\,=\,\,-\,\, \sum^{\infty}_{n,\,m=1}\,\,(-1)^n\,
\int\,\,\rho^t_m(\bar r_1, \,\bar b_1\ldots \,,\bar r_n,\, \bar b_n;\,Y_0)\,\,
\rho^p_n(r_1, \, b_1\,\ldots\,,r_n,\, b_n;\, Y\,-\,Y_0) \,\,\,\,\,\,\eeq
$$ \times \,\,\si^{n\,m}(r_1,\, b_1,\dots r_n,\,b_n\, |\,
\bar r_1,\, \bar b_1,\,\dots,\,\bar r_m,\,\bar b_m)\,\,
\prod^n_{i =1}\, \,
d^2\, r_i \, d^2\, b_i 
\,\,\prod^m_{j =1}\, \,d^2 \,\bar r_j \, d^2 \,\bar b_j\,.
$$
Eq. (\ref{AMPZ1}) appears   symmetric with respect to the exchange of the 
target and  projectile. In fact this symmetry is illusive. As soon as we 
require $\rho^p_n$ to obey the evolution given by Eq. (\ref{DFZ}) the boost 
invariance would impose a different evolution law 
(which we will not be able to write explicitly) for $\rho_m^t$.
We would like to emphasise the importance of the summation over $n$.
It guarantees the unitarity, and boost invariance of the amplitude. 
It is challenging to include simultaneously 
high dipole density effects, in both the  evolution of  
target and projectile.  Eq. (\ref{AMPZ1}) would then  account
for all the effects under discussion in Ref. \cite{PL}.

If $Y_0\,=\,0$ 
Eq. (\ref{AMPZ}) yields the expression for the amplitude in which
the whole evolution is through the projectile wavefunction. The dipoles
produced via the evolution  then multiple rescatter off the target
resulting in the unitarity respecting amplitude. The unitarisation is
achieved without saturation, as no high density effects are accounted by the
evolution of the projectile.

Alternatively we can take $Y_0\,=\,Y$. This would bring the whole evolution
into the target averaged amplitudes. 
By requirying the total amplitude $N$ to be independent
of our choice of frame we will be able to deduce the evolution law for
$\gamma_n(Y)$. As can be expected  $\gamma_n$ happens to obey the Balitsky's
chain of hierarchy equations \cite{B} projected onto dipole operators.

We now  derive the evolution of $\gamma_n$. We note that
$$
\rho^p_n\,=\,-\,(-1)^n\,\frac{\delta\,N}{\delta\,\gamma_n}
$$ 
We obtain the linear functional equation for the amplitude $N$:
$$
\frac{\partial N}{ 
\bar{\alpha}_s\,\partial\,Y}\,\,=\,\, \sum^{\infty}_{n=1}\,\,
\int\,\prod^n_{i =1}\,
d^2 \,r_i \, d^2\, b_i \,\,
\gamma_n(r_1, b_1,\ldots \,,r_n, b_n;\,Y_0)\,\,
\,\,\left(
-\,\sum_{i=1}^n
 \,\,\omega(r_i)\,\,\frac{\delta\,N}{\delta\,\gamma_n}\,\,+\right.
$$
\beq \label{DFN} 
\left.
2\,\sum_{i=1}^n\,
\int\,\frac{d^2\,r'}{2\,\pi}\,\,
\frac{r'^2}{r^2_i\,(r_i\,-\,r')^2}\,\,
\frac{\delta\,N}{\delta\,\gamma_n}\,\, 
-\,\,\sum_{i=1}^{n-1}\,\frac{(r_i + r_n)^2}
{(2\,\pi)\,r^2_i\,r^2_n}\,\,\frac{\delta\,N}{\delta\,\gamma_{n-1}}\,\right)\,.
\eeq
The equation (\ref{DFN}) can be solved using the ansatz:
$N(Y,[\gamma])=\,N(\gamma_1(Y), \gamma_2(Y)\, \ldots )$ such that
$$
\frac{\partial N}{\partial\,Y}\,=\, \sum^{\infty}_{n=1} \,\int\,\,
\frac{\delta\,N}{\delta\,\gamma_n}\,\,\frac{\partial\,\gamma_n}{\partial\,Y}\,
\,\,\prod^n_{i =1}\,
d^2\, r_i\, \, d^2\, b_i \,.
$$
As all $\gamma_n$ are independent of each other,  Eq. (\ref{DFN}) is equivalent
to Balitsky's hierarchy (restricted to dipole operators): 
$$
\frac{\partial \gamma_n(r_1, \,b_1,\ldots \,,r_n,\, b_n; \,Y)}{ 
\bar{\alpha}_s\,\partial\,Y}\,\,=\,\,
-\,\,\sum_{i=1}^n\,
\,\omega(r_i)\,\, \gamma_n(r_1,\, b_1,\ldots \,,r_n,\, b_n)\,\,+
$$
\beq \label{DFN1} 
\,\sum_{i=1}^n\,
\int\,\frac{d^2\,r'}{2\,\pi}\,\,
\frac{r_i^2}{r'^2\,(r_i\,-\,r')^2}\,\,\left[
2\,\gamma_n(r_1, \,b_1,\ldots \,r',\, b_i \,+\, \frac{(r_i\,-\,r')}{2},\,
\ldots\,,r_n,\, b_n)\,\, \,\,\,\,\,\,\,\,\,\,\right.
\eeq
$$
\left. \,\,\,\,\,\,\,\,\,\,\,\,\,\,\,\,\,\,\,\,\,\,\,
- \,\, \gamma_{n+1}(r_1, \,b_1,\ldots \,r', \,b_i\, +\, \frac{(r_i\,-\,r')}{2},
\,\ldots\,(r_i \,-\, r'), \,b_i\, +\,\frac{r'}{2})\,\right]\,.
$$
The initial conditions for the evolution chain (\ref{DFN1}) are given
by a target model which is supposed to provide $\gamma_n (Y\,=\,Y_0)$.
All $\gamma_n$ are defined as scattering amplitudes and hence are target 
averaged quantities.  In principal, 
Eq. (\ref{DFN1}) governs the evolution of the target densities $\rho_n^t$.
Though we are not able to write an explicit equation for $\rho_n^t$, it
apparently differs from the evolution of $\rho_n^p$ (Eq. (\ref{DFZ})). 

The amplitude $N$ can be rewritten in the following form 
$$
 N(Y) \,\,=\,\, -\,\,\sum^{\infty}_{n=1}\,(-1)^n\,
\int\,\gamma_n(r_1,\,b_1,\ldots \,,r_n, \,b_n;\,Y\,-\,Y_0)\,\,
\rho^p_n(r_1,\, b_1,\,\ldots\,r_n,\,b_n\,;\,Y_0)
\,\,\prod^n_{i =1}\,
d^2\, r_i\, \, d^2\, b_i \,.
$$
The $Y$ evolution has been transferred  from the evolution
in the projectile wavefunction ($\rho^p_n$)  to the evolution of 
the target averaged amplitudes ($\gamma_n$). Note  that by construction $N(Y)$ is boost or frame
invariant.  In other words, Eq. (\ref{DFN1}) can be obtained just by requirying
$N(Y)$ to be $Y_0$ independent. 
The functions $\rho^p_n(Y\,=\,Y_0)$ have to be provided by the initial dipole
distribution in the projectile. 
So far, our treatment was  general. Note, however, the asymmetry
between the target and projectile, as for the latter no high density effects
are included. This asymmetry is reflected by the fact that $\rho_n^p$
and $\gamma_n$ obey different hierarchy equations.

We would like to emphasise that so far our discussion was quite general
and valid for any projectile. Our results can be trusted until we violate
our central assumption, namely the  absence of nonlinear effects in the projectile
wave function.

We now make a choice of a projectile.
Consider the canonical case of a dipole of the size $r$ scattering
off a generic target at the impact parameter $b$. 
Then the initial condition in the projectile wave function
 at $Y\,=\,Y_0$ is
 \quad
$\rho^p_1\,=\,P_{n=1}\, =\, 
\delta^2( r\,-\,r_1)\,\delta^2( b\,-\,b_1)$ 
while $\rho^p_{n>1}\,=\,P_{n>1}\,=\,0$. 
This choice of the projectile 
initial conditions implies
\beq \label{LDIN1}
Z\left(Y \,=\,Y_0,\,r,\,b;\,[u]\right)\,\,=\,\,u(r, b)\,\,.
\eeq
For the  total amplitude $N$ we immediately obtain 
$$N(Y, \,r,\, b)\,=\,\gamma_1(r, \,b;\,Y)
$$
For a generic target we cannot write down a closed form equation
for $N$. 
It was shown in Ref. \cite{LL}  that for this particular 
choice of the projectile, 
\eq{LEQZF}  can be rewritten in the nonfunctional but
non-linear form, reproducing the same equation for $Z$ 
as in Ref. \cite{MUUN}:
\beq \label{NLEQZF}
\,\frac{\partial\, Z\left(Y, \,r,\, b;\,[u] \right) }{
\bar{\alpha}_s\,\partial \,Y}\,\,=\,\,- \,\,\omega(r)\,\,
Z\left(Y; \,r,\, b;\,[u] \right)
\eeq
$$
+\,\,\int\,\,
\frac{d^2\,r'}{2\,\pi}\,\,\frac{r^2}{r'^2\,(r\,-\,r')^2}\,\,
Z\left(Y;\,r',\, b\,+\,\frac{(r\,-\,r')}{2};\,[u] \right)\,\,
Z\left(Y;\,(r\,-\,r'),\, b\,-\,\frac{r'}{2};\,[u] \right)\,.
$$
The linear functional equation (\ref{LEQZF}) is more general, however, 
as it is valid for any projectile.

\section{Target correlations} 
 
Target correlations are defined  by having 
 $$\gamma_n(r_1,\,b_1,\ldots\,r_n,\,b_n;\,Y=Y_0)\,=\,
C_n(r_1,\,b_1\ldots\,r_n, \,b_n)\,
\,\gamma(r_1,\,b_1)\,\ldots\,\gamma(r_n,\,b_n)$$
 The coefficients $C_n$
are $n$-dipole correlation parameters. They can be also viewed as
effective measures of target fluctuations.
If we assume $C_n$  be
pure numbers independent of  coordinates, then Eq. (\ref{DFN})
is reduced to the following functional equation \cite{LL}
\beq \label{AMPLEQZ}
\frac{\partial\, N(Y;\,[\gamma])}{\bar \alpha_s\,\partial\,\,Y}  
\,\,=\,\,- \int\,d^2\,r' \,\,V_{1\rightarrow 1}(r',\,[\gamma(r')])\,\,
N(Y;\,[\gamma])\,\,
\eeq
$$
+\,\,\frac{2}{2\,\pi}\,\int \, d^2 \,r'\,d^2 \,r" \,
\,\gamma(r")\,\frac{r'^2}{\,r"^2\,( r"\,-\, r')^2}\,\,
 \frac{\delta}{\delta\, \gamma(r')}
\,\,N(Y;\,[\gamma ])
$$
$$
+\,\,F\left(\int \,d^2\,\bar r 
\,\gamma(\bar r)\,\frac{\delta}{\delta  
\gamma(\bar r)}\right)\,\,
\int \,d^2\,r"\,\,d^2\,r'
 \,\,V_{1\rightarrow 2}(r',\,r",\,[\gamma(r')])\,\, 
N(Y;\,[\gamma ])
$$
Here $F(n)\,\equiv C_n/C_{n-1}$. If we further simplify the correlations
assuming $F$ is constant
\beq\label{Cn}
C_n\,\,=\,\,F^{n\,-\,1}\,\,C_1
\eeq
then
\eq{AMPLEQZ} can be easily reduced to a non-functional non-linear equation  
(ignoring the $b$ dependence):
\beq    \label{NLEQRA}
\frac{\partial\,N(r,\,Y)}{\partial\,Y}\,=\,
\bar\as\,\times
\int_{\rho^p} \, \frac{d^2\,r'}{2\,\pi}\,
\frac{r^2}{r'^2\,(r\,-\,r')^2}\,\,\times
\eeq
$$
\left[\,2\,N(r',\,Y)
 \,\,-\,\,N(r,\,Y)\,\,-\,\,F/C_1\,\,N(r',\,Y)\,\,
N(r\,-\, r',\,Y)  
\right]\,, $$
The initial condition is $N(Y\,=\,Y_0)\,=\,C_1\,\,\gamma$.
Note that unless target properties are  specified,  $F$ and $C_1$ 
are arbitrary numbers ($F\,\ge\,C_1$). 

Our solution first obtained
in Ref. \cite{LL},  proves that under
condition (\ref{Cn}) the hierarchy  (\ref{DFN1}) shrinks to a single 
nonlinear equation (\ref{NLEQRA}). An identical result has been 
obtained recently in Ref. \cite{Janik} by a direct analysis of the hierarchy equations,
whereas the result $F/C_1=2$ of Ref. \cite{JP} is a particular solution.

The new aspect which we find as most important,  is the relation 
between the target correlations and Balitsky`s hierarchy, something which 
we were not aware of when preparing Ref. \cite{LL}. 
Eq. (\ref{NLEQRA}) which we derived was 
viewed  as a phenomenologically  motivated ad hoc modification 
of the BK equation.  By  establishing in this letter a relation between the 
correlations and  hierarchy,  we attempt to correct the wrong attitude 
regarding  our  original work.

If we consider dipole interaction as fully uncorrelated $F\,=\,C_1\,=\,1$
then \eq{NLEQRA} reduces to the BK equation. 
Its solution can  also be  written
in the following form
$$N(Y,\,r,\,b)\,=\,-\,\,\sum^{\infty}_{n=1}\,(-1)^n\,
\rho^p_n(r_1,\, b_1,\,\ldots\,r_n,\,b_n\,;\,Y\,-\,Y_0)
\,\,\prod^n_{i =1}\,N(Y_0,\,r_i,\,b_i)\,\,
d^2\, r_i\, \, d^2\, b_i \,.
$$ 

So far the approximation $F\,=\,const$ looks pure mathematical. 
In Ref. \cite{LL}, a realistic model for target correlations was proposed.
In that model the target was considered to be a nucleus (not necessary
very heavy) but the approach can be viewed more generally.
The target correlations  were  estimated  based on pure counting 
arguments  independent of the dipole coordinates. 
The  model of Ref \cite{LL} leads to a constant $F$,
providing a physically intuitive
realisation of a more formal structure outlined above.

\section*{Acknowledgments}

We are most grateful to Edmond Iancu who in fact suggested and inspired
us to derive the results reported in this paper.
\\
We wish to thank  Errol Gotsman and Alex Kovner  for most fruitful discussions which went
far beyond this paper.
\\
This work was initiated when M.L. was visiting
SPhT of Saclay. M.L. would like to express his gratitude to
SPhT for the warm hospitality. Stimulating discussions with Robi
Peschanski and other members of SPhT are thankfully acknowledged.

\end{document}